\documentclass[twocolumn,english,preprintnumbers,amsmath,amssymb,prb]{revtex4}

\usepackage{graphicx}
\usepackage{color}

\begin{document}

\title{Cubic silicon carbide under tensile pressure: Spinodal instability}
\author{Carlos P. Herrero}
\author{Rafael Ram\'irez}
\affiliation{Instituto de Ciencia de Materiales de Madrid,
         Consejo Superior de Investigaciones Cient\'ificas (CSIC),
         Campus de Cantoblanco, 28049 Madrid, Spain}
\author{Gabriela Herrero-Saboya}	 
\affiliation{CNR-IOM Democritos National Simulation Center, 
         Istituto Officina dei Materiali, c/o SISSA, 
	 via Bonomea 265, IT-34136 Trieste, Italy}

\date{\today}

\begin{abstract}
Silicon carbide is a hard, semiconducting material presenting
many polytypes, whose behavior under extreme conditions of
pressure and temperature has attracted large interest.
Here we study the mechanical properties of $3C$-SiC over a wide 
range of pressures (compressive and tensile) by means of molecular  
dynamics simulations, using an effective tight-binding Hamiltonian
to describe the interatomic interactions. The accuracy of this
procedure has been checked by comparing results at $T = 0$
with those derived from {\em ab-initio} density-functional-theory
calculations.
This has allowed us to determine the metastability limits of
this material and in particular the spinodal point 
(where the bulk modulus vanishes) as a function of temperature.
At $T= 300$~K, the spinodal instability appears for a lattice 
parameter about 20\% larger than that corresponding to ambient 
pressure.  At this temperature, we find a spinodal pressure 
$P_s = -43$~GPa, which becomes less negative as temperature 
is raised ($P_s = -37.9$~GPa at 1500 K).
These results pave the way for a deeper understanding of the
behavior of crystalline semiconductors in a poorly known
region of their phase diagrams.   \\

\end{abstract}

\maketitle

\section{Introduction}

In the last few decades, the experimentally accessible region
of the phase diagrams of different substances has been expanded,
giving us deeper understanding of condensed phases under extreme
conditions of pressure and temperature.\cite{sc-mu03,sc-ma18}
Thus, the effect of hydrostatic pressure on various properties of
different types of solids have been largely analyzed.
This includes a  growing interest in condensed matter under
tensile stress, which can yield information on the metastability
limits of different phases as well as about the attractive region of
interatomic potentials.\cite{sc-da10,sc-iy14,sc-ni19,sc-im08c}

The behavior of condensed matter under hydrostatic tensile pressure 
has been mainly explored for liquids.\cite{sc-so92,bo94b,
sc-je03,sc-im07,sc-im08,sc-da10}  This has included the study of 
the limits for mechanical stability, with the appearance of 
cavitation close to the corresponding spinodal lines. 
This kind of phenomena have been also studied for various types 
of solids, so that extreme pressure conditions do not only refer
to large compressive stress, but also to tensile stress (negative
pressure).\cite{sc-th85,he03b,sc-iy14,sc-pe15,sc-li18,sc-ni19,sc-ve22}
It has been shown that tensile pressure can be relevant to understand 
unexplored regions of stability of solids under hydrostatic 
(or quasi-hydrostatic) conditions. 
In this context, one aim of the present paper is to gain insight
into the spinodal lines of semiconducting crystalline solids,
which delineate the limit of mechanical stability of these materials.
These lines are still poorly understood for well-known solids
such as Si or silicon carbide. In particular, we concentrate
here on cubic $3C$-SiC with zinc-blende-type structure 
(also called $\beta$ or $B3$ phase).

Silicon carbide under compressive hydrostatic pressure, including
phase transitions, has been studied in detail earlier both
theoretically\cite{sc-sh00,ra08,sc-va15,sc-le15b,sc-ra21,sc-da22,sc-pe22}
and experimentally.\cite{sc-zh13,sc-ni17,sc-da17,sc-da18,sc-mi18,sc-ki22}
The interest in the high-pressure behavior of semiconducting materials
has recently risen, apart from the traditional context of condensed
matter physics, as potential constituents of carbon-rich exoplanets.
Various studies have found that high pressure in planetary interiors
may significantly change the physical properties of these
materials.\cite{sc-ni17,sc-mi18,sc-ki22}

Given the large amount of work carried out for $3C$-SiC under
compressive pressure, 
we focus here mainly on the effects of tensile stress.
In this context, crystalline silicon has been investigated at negative 
pressure by molecular dynamics simulations, yielding information 
on the appearance of cavitation and crystal-liquid 
interfaces,\cite{sc-wi03,sc-da10b} as well as on the transition from 
diamond-type structure to a clathrate at a pressure 
$P \approx -2.5$~GPa.\cite{sc-ka05}

The experimentally reachable range of hydrostatic (or quasi-hydrostatic)
tensile pressure has been growing along the years, as well as
the understanding of physical properties under conditions hardly 
accessible in the laboratory.\cite{sc-da10,sc-he87,sc-gr88,sc-mo00,
sc-du02,sc-ve22} 
Detailed quantitative experimental studies of materials under negative 
pressure are scarce as one works in metastable conditions, which in many
cases are only available for short periods of time.
In our present context, carbon-based materials have been studied 
under tensile pressure by means of ultrasonic cavitation and shock waves 
created by picosecond laser pulses.\cite{sc-ab14b,sc-kh08}
Moreover, low-density clathrates, allotropes of group IVa elements 
(C, Si, Ge), which turn out to be metastable at ambient conditions, 
have been synthesized in recent years.\cite{sc-ba14,sc-gu06}

In this paper we explore the metastability region of
cubic SiC under tensile hydrostatic pressure. We present results
of molecular dynamics (MD) simulations carried out using
interatomic interactions based on a reliable tight-binding (TB)
Hamiltonian. Density-functional-theory (DFT) calculations 
were performed at $T = 0$ to assess the precision of the TB results 
for conditions (crystal volume) far from equilibrium.
We find that $3C$-SiC is metastable in a wide pressure range
till $P \sim -40$~GPa, and MD simulations allow us to approach 
the limit of mechanical stability of the solid (spinodal pressure), 
which is obtained at temperatures up to 1500~K.

The paper is organized as follows. In Sec.~II we present the
computational methods used in the calculations.
In Sec.~III we show the phonon dispersion bands and the
elastic constants of $3C$-SiC.  Results for the energy obtained 
from TB and DFT methods are given in Sec.~IV.
The spinodal instability appearing at negative pressures is
discussed in Sec.~V, and a summary of the main results is presented
in Sec.~VI.

\section{Computational method}

In this section we present the methods used in this paper.
In Sec.~II.A we concentrate on molecular dynamics simulations and the
tight-binding procedure employed to define the interatomic
interactions.
In Sec.~II.B, we introduce the DFT-based approach employed to
evaluate the accuracy of the TB results at $T = 0$.

\subsection{Tight-binding molecular dynamics}

We investigate structural and mechanical properties of $3C$-SiC 
as functions of temperature and pressure using MD simulations.
A relevant point in the MD method is the consideration of realistic 
interatomic interactions, which should be as reliable as possible.
One could achieve this goal by employing {\em ab-initio} density 
functional or Hartree-Fock based self-consistent potentials for
finite-temperature simulations, but this would enormously limit
the length of the simulation trajectories which could be obtained
in a reasonable computing time.
We then determine the interatomic forces from an effective
tight-binding Hamiltonian, built up from results of density functional 
calculations.\cite{po95}

This type of TB methods display good accuracy to describe several
properties of condensed matter and molecular systems, as discussed by
Goringe {\em et al.}\cite{go97} and Colombo.\cite{co05}
The TB Hamiltonian used here\cite{po95} was found before to be 
trustworthy to define the interatomic  interactions in
carbon-based materials.\cite{he06,he07}
The parametrization for structures containing Si and C  atoms
was given in Ref.~\onlinecite{gu96}.
The non-orthogonality of the atomic basis is a crucial clue for 
the transferability of the parametrization to complex systems \cite{po95}.

The tight-binding method was employed earlier to study silicon
carbide, especially the cubic phase considered
here.\cite{sc-me96,sc-be05}
In particular, the TB Hamiltonian used in this work has been
applied before to study reconstructions of $3C$-SiC
surfaces,\cite{gu96,sc-sh01} as well as to investigate isotopic
and nuclear quantum effects in this material.\cite{ra08,he09c}
In relation with our present work, it was
employed to study this crystalline solid under compressive
pressure up to 60~GPa, with results that compared well with
experimental data and other theoretical calculations.\cite{ra08}
More recently, this TB model has been applied to analyze several 
properties of the lately synthesized monolayers of silicon 
carbide.\cite{he22,sc-po23}

TB MD simulations have been carried out in the
isothermal-isobaric ($NPT$) ensemble for cubic SiC supercells
including 64 and 216 atoms. Some simulations were performed
in the canonical ($NVT$) ensemble near the limit of mechanical 
stability, as they allow to perform simulations closer to the 
spinodal point, in a region where $NPT$ simulations are unstable 
due to the appearance of large volume fluctuations.
Periodic boundary conditions were assumed in all cases.

To keep the required temperature, chains of four Nos\'e-Hoover
thermostats were coupled to each atomic degree of freedom.\cite{tu98}
For $NPT$ simulations, an additional chain of four thermostats 
was connected to the barostat that controls the volume of the 
simulation cell, giving a constant pressure.\cite{tu98,al87}
The equations of motion were integrated by employing the reversible
reference system propagator algorithm (RESPA), which permits
to use different time steps for the integration of slow and fast
degrees of freedom.\cite{ma96}
The time step $\Delta t$ employed for the dynamics associated to
the forces derived from the TB Hamiltonian was 1 fs, which gives 
good accuracy for the temperatures studied here.
For fast dynamical variables such as the thermostats, we used
a time step $\delta t = 0.25$ fs.
The configuration space has been sampled for temperatures 
from 300 to 1500 K.
Given a temperature, a typical run consisted of $2 \times 10^5$ MD
steps for system equilibration, and  $8 \times 10^6$ steps
for the calculation of mean variables.

For sampling electronic degrees of freedom in the reciprocal space,
we have considered only the $\Gamma$ point (${\bf k} = 0$).
The main consequence of employing larger ${\bf k}$ sets is 
a shift of the total energy, with imperceptible change in energy 
differences between various atomic configurations.
Something similar occurs for the mean energy per atom 
at a given temperature, for different cell sizes.\cite{he22}
This has been checked here for silicon carbide using simulation 
cells of size $N$ = 64 and 216 atoms.

We calculate the elastic constants of cubic SiC at finite 
temperatures by applying a particular component of the stress 
tensor $\{ \tau_{ij} \}$ in isothermal-isobaric simulations, and 
finding the compliance constants $S_{ij}$ from the obtained strain.
For instance, for $\tau_{xx} \neq 0$ and $\tau_{ij} = 0$
for the other components, we have
$S_{11} = e_{xx} / \tau_{xx}$, and $S_{12} = e_{yy} / \tau_{xx}$,
$e_{ij}$ being the components of the strain tensor.\cite{as76,ki05,yu96}
From the compliance constants, we obtain the stiffness constants 
$C_{11}$ and $C_{12}$ by means of the relations corresponding to 
cubic crystals:\cite{as76,ki05} $C_{11} = (S_{11} + S_{12}) / Z$ 
and $C_{12} = - S_{12} / Z$, with
\begin{equation}
     Z = (S_{11} - S_{12}) (S_{11} + 2 S_{12})   \; .
\end{equation}
Note that a hydrostatic pressure $P$ corresponds in the elasticity 
notation to $\tau_{xx} =  \tau_{yy} = \tau_{zz} = -P$.
Here, $P > 0$ and $P < 0$ represent compressive and tensile 
pressure, respectively.

\subsection{DFT calculations}

In order to verify the accuracy of our TB method for describing
the mechanical properties of cubic SiC, we have performed
state-of-the-art DFT calculations.
Total energies were obtained with the Quantum-ESPRESSO
package for electronic structure calculations.\cite{sc-gi09,sc-gi17} 
We employed the Perdew-Burke-Ernzerhof exchange-correlation
functional, as adapted for solids (PBEsol),\cite{sc-pe08} using a 
plane wave basis set with a kinetic energy cutoff of 45 Ry (400 Ry 
for the charge density cutoff). Projector-augmented-wave (PAW)
pseudopotentials were employed for both carbon and silicon.\cite{sc-ps23}
We considered a cubic cell of SiC containing 8 atoms 
with zinc-blende structure, subject to periodic boundary
conditions. The Brillouin zone was integrated with a 
$10 \times 10 \times10$ Monkhorst-Pack grid.\cite{sc-mo76}

{\em Ab-initio} calculations were carried out earlier to study 
several aspects of $3C$-SiC, such as structural, electronic, 
elastic, lattice-dynamical, and thermodynamic 
properties.\cite{sc-ch86,sc-pa94b,sc-ka94,sc-ka94b,sc-ca20}    
In the context of our present work, they have been employed to
characterize phase transitions in this material under high
pressure.\cite{sc-ch87,sc-le15b,sc-ki17,sc-sh18,sc-ra21}

\section{Energy}

In Fig.~1 we present the energy per atom as a function of the 
lattice parameter $a$ of $3C$ silicon carbide.
The solid line represents results of PBEsol-PAW DFT calculations.
We find for the minimum-energy configuration a lattice parameter
$a_0$ = 4.358~\AA, in agreement with the result of 
Lee and Yao.\cite{sc-le15b}
Note that the zero of energy is taken at $a_0$.
The dashed line corresponds to our tight-binding calculations
at $T = 0$. It displays a minimum at $a_0$ = 4.346~\AA, and follows 
closely the DFT result for lattice parameters around $a_0$.
For large values of $a$,
in the region where the material becomes unstable, both lines 
progressively depart from one another, the TB energy being
higher than that corresponding to the DFT calculations, and 
for $a =$~5.5 \AA\ the difference between both amounts to 0.17 eV.

\begin{figure}
\vspace{-0.0cm}
\includegraphics[width=7cm]{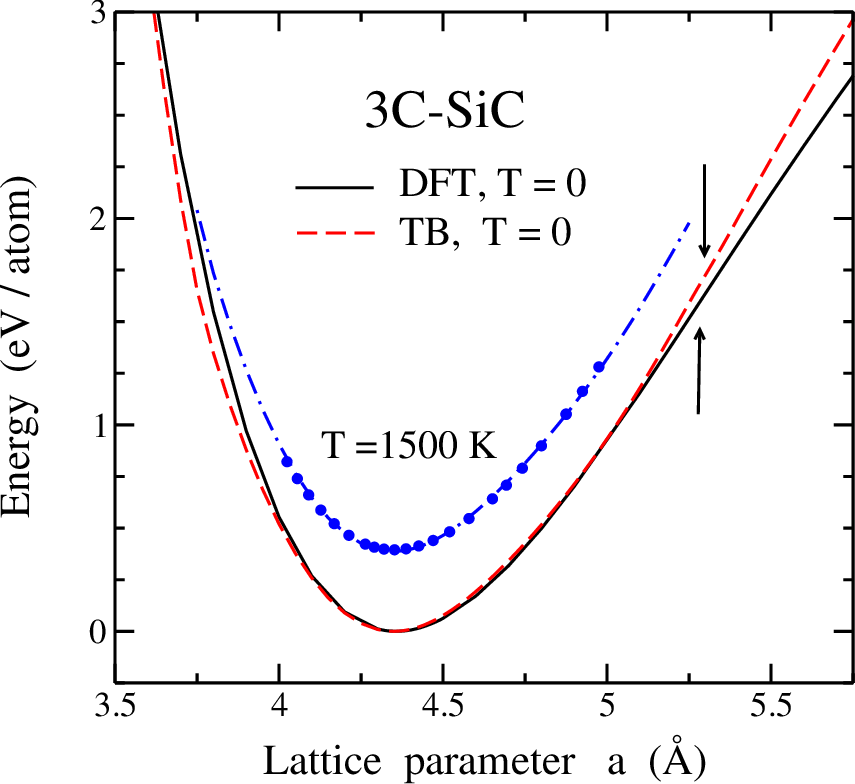}
\vspace{0.3cm}
\caption{Energy vs lattice parameter $a$ of $3C$ silicon
carbide. The solid and dashed lines represent results of
DFT and TB calculations at $T = 0$. Solid circles indicate
results of MD simulations at $T = 1500$~K.
The dashed-dotted line corresponds to an upwards shift of the
$T = 0$ TB line by an amount of $3 k_B T$ for 1500 K.
Error bars of the simulation data are less than the symbol size.
Two vertical arrows show the inflection points of the solid
and dashed curves.
}
\label{f1}
\end{figure}

At $T = 0$, the hydrostatic pressure is given by 
$P = - \partial E / \partial V$.
Thus, for the TB model, the pressure $P$ corresponding to lattice 
parameter $a$ from 3.7 to 5.25 \AA\ goes from 303 to $-44$~GPa.
Note that the range of lattice parameters that are explored
with compressive pressure ($P > 0$) up to about 300~GPa
corresponds to a reduction of $a$ by a 15\%.
On the contrary, tensile pressure causes expansion of 
the lattice with an increase in $a$ by a 21\% up to
the stability limit for $P \approx -44$~GPa.

To show the effect of temperature, we also
display in Fig.~1 results of MD simulations at $T = 1500$~K
(solid circles). These data were obtained from simulations
in the isothermal-isobaric $NPT$ ensemble for hydrostatic
pressures between 80~GPa ($a$ = 4.02~\AA) and 
$-37.3$~GPa ($a$ = 4.97~\AA).
Note that the latter pressure is near the spinodal
pressure $P_s$, where the solid becomes unstable at 
$T = 1500$~K (see below).
At this pressure and temperature we find in the MD simulations
an energy $E$ = 1.28~eV/atom. This energy can be split into a
contribution of 0.89~eV/atom due to elastic energy (lattice
expansion) and another of 0.39~eV/atom due to thermal energy,
$E_{\rm th}$, at this temperature. 
This means that at this relatively high temperature,
$E_{\rm th}$ represents a 30\% of the total energy close
to the spinodal pressure $P_s$.

For comparison with the simulation results at $T$ = 1500~K, 
we present in Fig.~1 the expected energy for a classical harmonic 
model for the lattice vibrations at each crystal volume
at this temperature (dashed-dotted curve). This is
obtained by adding an energy of $3 k_B T$ ($k_B$, Boltzmann's 
constant) to the zero-temperature TB result. 
One observes that both finite-temperature data sets follow 
each other closely.

For our later discussion on the mechanical stability of
$3C$-SiC, it is interesting to determine the inflection point
of the curves displayed in Fig.~1. This point separates the
regions where they are concave upward ($d^2 E / d a^2 > 0$) and
downward ($d^2 E / d a^2 < 0$), and is represented by vertical
arrows for the $T = 0$ curves in Fig.~1.

\begin{figure}
\vspace{-0.0cm}
\includegraphics[width=7cm]{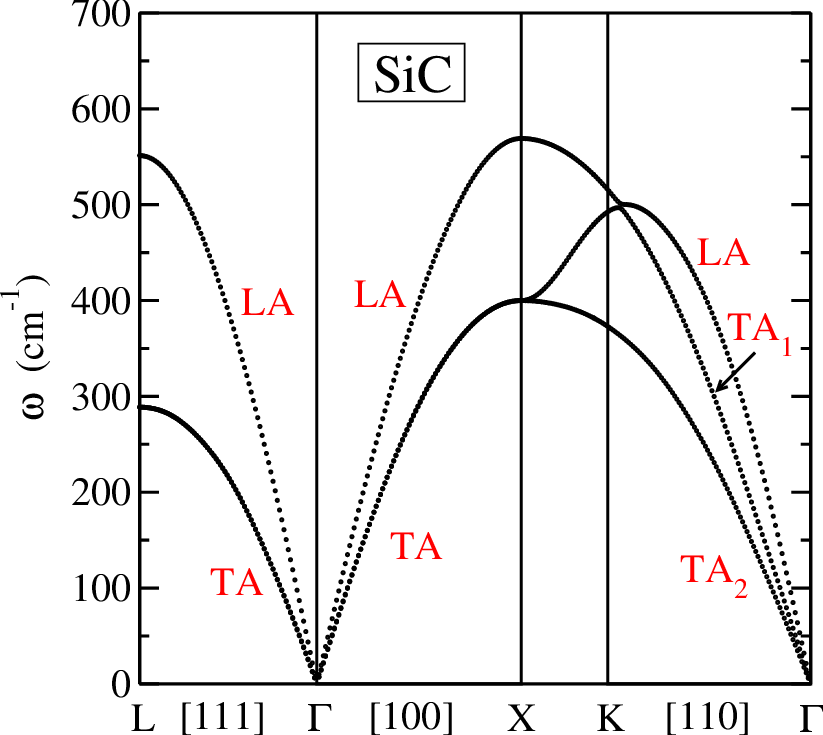}
\vspace{0.8cm}
\caption{Acoustic phonon bands of $3C$-SiC derived from the
dynamical matrix corresponding to the tight-binding model
for the minimum-energy volume.
Labels indicate the character of the different branches:
LA, longitudinal acoustic; TA, transversal acoustic.
The TA band along $[111]$ and $[100]$ directions is
twofold degenerate.
}
\label{f2}
\end{figure}

\section{Phonon dispersion bands and elastic constants}

The elastic stiffness constants $C_{ij}$ of cubic SiC calculated
with the TB Hamiltonian for $T = 0$ may be taken as
a reference for the finite-temperature analysis presented below. 
We obtain these elastic constants in the low-$T$ limit from 
the harmonic dispersion relation of acoustic phonons.
To define the dynamical matrix necessary to find the phonon bands, 
we calculated the interatomic force constants by numerical 
differentiation of atomic forces, taking atom displacements 
of $2 \times10^{-4}$~\AA\ from the minimum-energy sites.
Good numerical convergence in the phonon bands was achieved 
by calculating all interatomic force constants up to 
distances of about 18~\AA.
In Fig.~2 we present the acoustic phonon branches obtained 
in this way for the minimum-energy configuration ($a_0 = 4.346$~\AA),
along symmetry directions of the Brillouin zone.
The phonon dispersion displayed in this plot is similar to those
found for other effective potentials and DFT
calculations,\cite{sc-ka94,sc-ta17,sc-wa17b,ba21}
and to the acoustic phonon bands obtained from inelastic 
x-ray scattering.\cite{sc-se02}

The sound velocities for the acoustic bands along the directions 
shown in Fig.~2 are given by the slope of the bands at the
$\Gamma$ point ($k \to 0$).
Here $k$ denotes the wavenumber, i.e., $k = |{\bf k}|$, and
${\bf k} = (k_x, k_y, k_z)$ is a wavevector in the Brillouin zone.
The elastic constants $C_{11}$ and $C_{12}$, relevant for our
discussion on the bulk modulus and the mechanical stability of
the solid, can be calculated from the expressions:\cite{ki05,yu96}
\begin{equation}
  C_{11} = \rho \left( \frac{\partial \omega_{\rm LA}}{\partial k_x }
           \right)^2_{\Gamma}  \; ,
\label{c11}
\end{equation}
for the LA band along the $[100]$ direction, and
\begin{equation}
  C_{12} = C_{11} - 2 \rho \left( \frac{\partial \omega_{\rm TA_2}}
           {\partial k}  \right)^2_{\Gamma}  \; .
\label{c12}
\end{equation}
for the TA$_2$ band along the $[110]$ direction. Here $\rho$ is
the density of the solid.
From the phonon bands shown in Fig.~2, using Eqs.~(\ref{c11}) and 
(\ref{c12}), we find $C_{11}$ = 452.9 GPa and $C_{12}$ = 141.1 GPa.
We have checked the consistency of these values with those obtained
from the slopes of the different bands at the $\Gamma$ point along 
the directions in ${\bf k}$-space shown in this figure.

\begin{figure}
\vspace{-0.0cm}
\includegraphics[width=7cm]{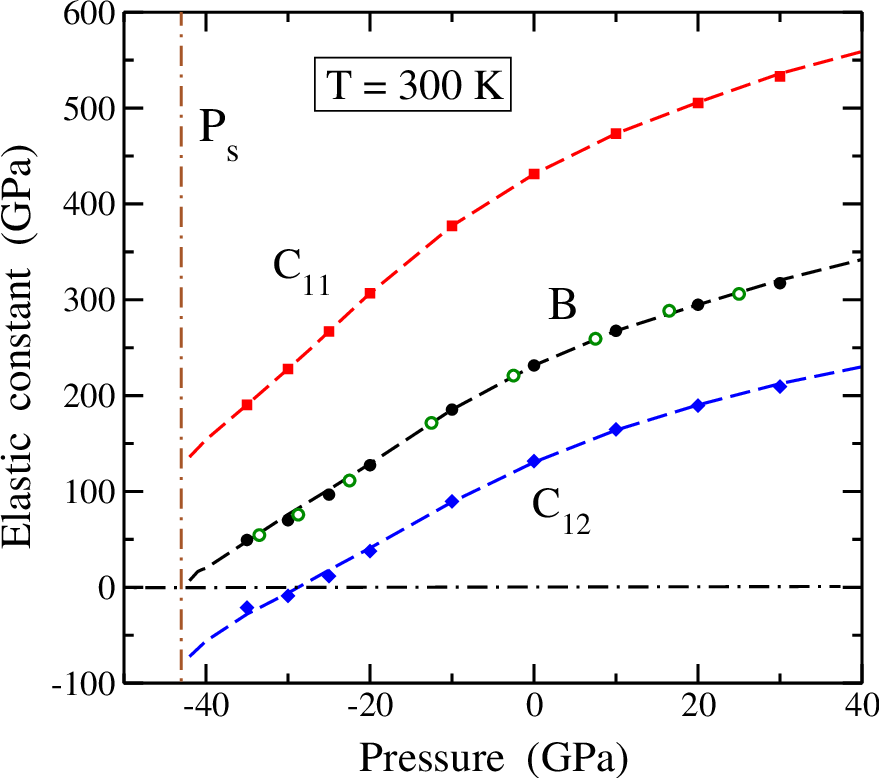}
\vspace{0.3cm}
\caption{Pressure dependence of the elastic constants $C_{11}$
(squares) and $C_{12}$ (diamonds) derived from MD simulations of
$3C$-SiC at $T =$ 300~K. Solid circles represent the bulk modulus
$B$ obtained from the elastic constants by using Eq.~(\ref{bulkm}).
Open circles indicate values of $B$ found from numerical
differentiation of the $P-V$ curve.
Error bars are in the order of the symbol size.
The vertical dashed-dotted line shows the spinodal pressure at 300~K.
Dashed lines are guides to the eye.
}
\label{f3}
\end{figure}

At finite temperatures, we have calculated the stiffness constants 
$C_{11}$ and $C_{12}$ from MD simulations as indicated above in 
Sec.~II.A.  For stress-free silicon carbide, we find an appreciable  
decrease in both elastic constants for rising temperature.
Thus, at $T$ = 300~K we have $C_{11} =$ 434.8 GPa and
$C_{12} =$ 129.4 GPa, which means a reduction of 4\% and 8\%, 
respectively, with respect to the zero-temperature values.
At the highest temperature considered here, $T = 1500$~K, 
we find a decrease of 15\% and 26\%, respectively, in comparison 
with the $T = 0$ values.

In Fig.~3 we show the dependence of $C_{11}$ and $C_{12}$
on hydrostatic pressure $P$ at 300~K. Symbols represent data
derived from our MD simulations. For positive (compressive)
$P$, we observe an increase of both elastic constants.
For increasing tensile (negative) pressure, the elastic constants
decrease, and at $P = -27$~GPa, $C_{12}$ becomes negative.
Note that $C_{11} > 0$ in the considered pressure range, as this
is a condition for mechanical stability of 
a solid phase.\cite{sc-ja14,mo14}

An important characteristic of solids concerning their elastic
properties is the Poisson's ratio $\nu$, which may be expressed
for a cubic phase as $\nu = C_{12} / C_{11}$. 
Thus we have in the classical
low-temperature limit $\nu$ = 0.31. This parameter changes for
rising temperature, as the elastic constants, and for ambient 
conditions ($T$ = 300~K, $P$ = 0) we find $\nu = 0.30$, close to
a value $\nu = 0.31(1)$ derived by Zhuravlev {\em et al.} from
x-ray diffraction and Brillouin spectroscopy.\cite{sc-zh13}
At 300~K, the Poisson's ratio yielded by our MD simulations
becomes negative as $C_{12}$ for $P = -27$ GPa, and cubic SiC 
transforms into an auxetic solid at this tensile pressure.

The isothermal bulk modulus, defined as
$B = - V \, \partial P / \partial V$,
can be obtained from the elastic constants by means of the 
expression, valid for cubic crystals:\cite{as76,ki05,sc-ja14}
\begin{equation}
   B = \frac13 (C_{11} + 2 C_{12})  \; .
\label{bulkm}
\end{equation}
From the elastic constants given above, we obtain at $T = 0$:
$B_0$ = 245.0 GPa.
We estimate an error bar of $\pm 2$~GPa, mainly caused by 
the uncertainty in the determination of the phonon band slopes
at the $\Gamma$ point.
The classical zero-temperature bulk modulus can be also obtained as
\begin{equation}
 B_0 = V_0 \left. \frac{\partial^2 E}{\partial V^2} \right|_{V_0} \; ,
\label{bulk0}
\end{equation}
where $E$ is the energy and $V_0$ is the volume for 
the minimum-energy configuration.
This gives for our TB Hamiltonian $B_0$ = 245.6~GPa, which agrees
with the value calculated from the elastic constants, taking into
account the error bars. 

From the elastic constants, we find (using Eq.~(\ref{bulkm}))
at $T = 300$~K and $P = 0$ a bulk modulus $B$ = 232(1)~GPa,
to be compared with experimental results\cite{sc-al89,sc-st87,sc-wa16} 
in the range from 224~GPa\cite{sc-ye71} to 260~GPa.\cite{sc-yo93}
Our value for 300~K means a reduction of about a 6\% with respect to 
the zero-temperature result. According to the data found for $C_{11}$ 
and $C_{12}$, we have for $T = 1500$~K a bulk modulus $B$ = 198(1)~GPa.
From the decrease in $B$ for rising $T$, we obtain around 
room temperature ($T$ = 300~K) a derivative 
$\partial B / \partial T = -0.040(2)$ GPa K$^{-1}$, 
close to the value derived by Wang {\em et al.}\cite{sc-wa16} 
from x-ray diffraction experiments: 
$\partial B / \partial T = -0.037(4)$ GPa K$^{-1}$.

The modulus $B$ is especially interesting to study the critical
behavior of silicon carbide under hydrostatic pressure.
In Fig.~3 we display, along with the elastic constants, the 
dependence of $B$ on $P$ at $T = 300$~K, including
tensile and compressive pressure. Solid circles indicate values 
of $B$ obtained from the elastic constants using Eq.~(\ref{bulkm}).
For comparison we also display as open circles results for $B$
obtained from numerical differentiation of the $P-V$ equation
of state at this temperature, using the expression 
$B = - V \, \partial P / \partial V$.  Results of both procedures 
agree well in the whole pressure region shown in Fig.~3 (error 
bars are in the order of the symbol size), which gives 
a consistency check for our calculations.

\section{Spinodal instability}

The material dilation due to tensile stress causes a fast decrease 
in the bulk modulus $B$, which vanishes for a pressure $P_s$, where
SiC becomes mechanically unstable.
This is typical of a spinodal point in the $(P, T)$ phase
diagram.\cite{sc95,he03b,ra18b,ca85}
For a given temperature $T$, there is a range of tensile pressure
where cubic SiC is metastable, i.e., for $0 > P > P_s$.
The spinodal line, which delineates the unstable phase ($P < P_s$) 
from the metastable phase, is the locus of points $P_s(T)$ where $B = 0$.
This type of spinodal lines have been investigated before
for water,\cite{sp82} as well as for ice, SiO$_2$ cristobalite,\cite{sc95}
and noble-gas solids\cite{he03b} close to their stability limits.
In the last few years, this question has been considered for
two-dimensional materials, in particular for graphene, where this kind 
of instability appears also for compressive stress.\cite{ra18b,ra20}

\subsection{Isothermal formulation}

Close to a spinodal point, the Helmholtz free energy for temperature 
$T$ can be written as a Taylor power expansion in terms of 
$V_s - V$:\cite{sp82,bo94b,ra20}
\begin{eqnarray}
    F(V, T) = F(V_s(T), T) + a_1(T) \, [V_s(T) - V] + \nonumber \\
	\hspace{-2cm}   a_3(T) \, [V_s(T) - V]^3 + ...  \; ,
\label{ffc}
\end{eqnarray}
where $V_s(T)$ and $F(V_s(T), T)$ are the volume and free energy 
at the spinodal point. At this point one has
$\partial^2 F / \partial V^2 = 0$, so that a quadratic term
does not appear on the r.h.s of Eq.~(\ref{ffc}), i.e., $a_2 = 0$.
Note that the coefficients $a_i$, as well as the spinodal volume 
$V_s$, are in general dependent on the temperature. 
In the following we will not write this dependence explicitly.

The pressure is
\begin{equation}
   P = - \frac{\partial F}{\partial V} =
           P_s + 3 a_3 \, (V_s - V)^2 + ...   \; ,
\label{pfv}
\end{equation}
and $P_s = a_1$ is the spinodal pressure, which corresponds
to the volume $V_s$.
The isothermal bulk modulus is given by
\begin{equation}
   B = V \, \frac{\partial^2 F}{\partial V^2} =
          - V \, \frac{\partial P}{\partial V}   \; ,
\label{bpa}
\end{equation}
and to leading order in an expansion in powers of $V_s - V$,
we have
\begin{equation}
   B = 6 a_3 V_s (V_s - V)   \; ,
\end{equation}
or considering Eq.~(\ref{pfv}), $B$ can be expressed along
an isotherm, close to the spinodal pressure $P_s$, as
\begin{equation}
   B = 2 \sqrt{3 a_3} \, V_s \, (P - P_s)^{1/2} \; .
\label{ba2}
\end{equation}
Thus, the bulk modulus vanishes for $P = P_s$, which gives the
limit of mechanical stability for the considered phase.

In the present work, most of the simulations have been carried out
in the isothermal-isobaric ensemble, and we determine $P_s$ as 
a function of $T$.
Similarly, for a given volume $V$, there are spinodal pressure $P_s(V)$ 
and temperature $T_s(V)$, and changing $P$ and $T$ along an isochore
close to the spinodal point, one has to first order the linear 
relation  $P - P_s(V) \propto  T - T_s(V)$.\cite{sp82}

\begin{figure}
\vspace{-0.0cm}
\includegraphics[width=7cm]{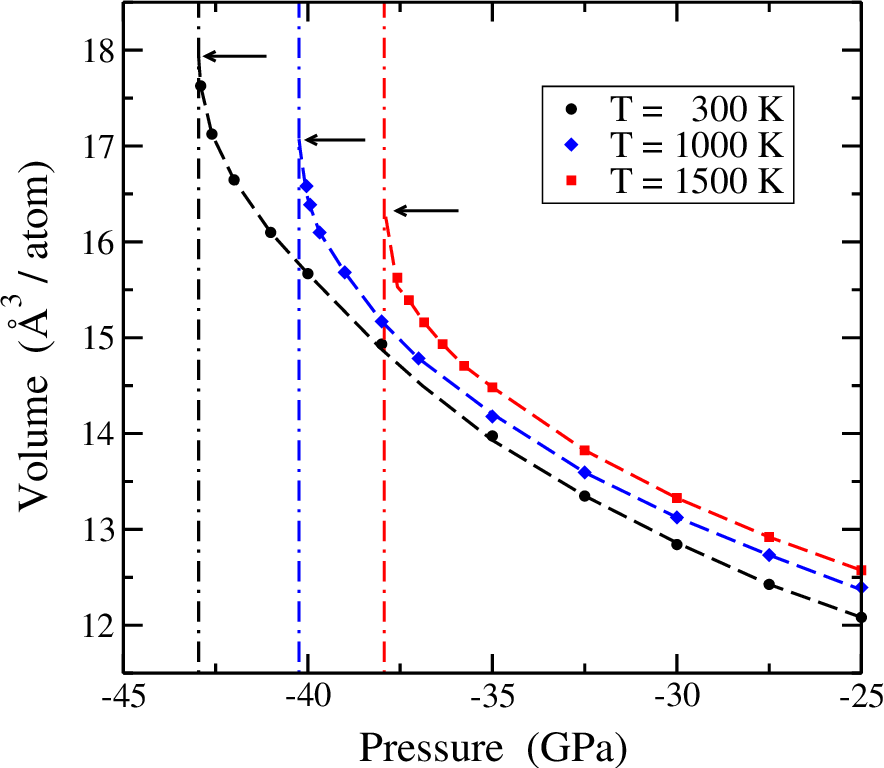}
\vspace{0.3cm}
\caption{Pressure dependence of the volume of cubic SiC,
as derived from MD simulations at $T$ = 300 K (circles),
1000 K (diamonds), and 1500 K (squares).
Vertical dashed-dotted lines indicate
the spinodal pressure for each temperature. Arrows show
the spinodal volume in each case.
Error bars of the simulation data are in the order of
the symbol size.
Dashed curves are guides to the eye.
}
\label{f4}
\end{figure}

\subsection{Application to $3C$-SiC}

In Fig.~4 we display the pressure dependence of the volume of
cubic SiC for $T$ = 300, 1000, and 1500 K. Solid symbols represent
results of MD simulations at various tensile pressures.
For each temperature, one observes an increase in volume for 
rising tensile pressure ($P$ more negative), i.e., $dV/dP < 0$,
as required for thermodynamic consistency. At a certain pressure
(spinodal) $dV/dP$ diverges to $-\infty$.
Note that the slope of each $P-V$ curve shown in Fig.~4 diverges 
for a spinodal volume $V_s(T)$, shown by a horizontal arrow.
The corresponding spinodal pressures $P_s(T)$ are indicated
by vertical dashed lines at $P_s = -43.0, -40.2$ and
$-37.9$~GPa for $T$ = 300, 1000, and 1500~K, respectively.

The large volume fluctuations appearing in $NPT$ simulations close
to the spinodal pressure do not allow us to reliably sample that region
of the configuration space. This limitation increases as the 
temperature is raised and the volume fluctuations also rise.
This problem is remedied in part by carrying out canonical ($NVT$)
simulations in those parts of the configuration space, where $3C$-SiC
remains metastable during simulation runs long enough to accurately 
sample the required thermodynamic variables.

\begin{figure}
\vspace{-0.0cm}
\includegraphics[width=7cm]{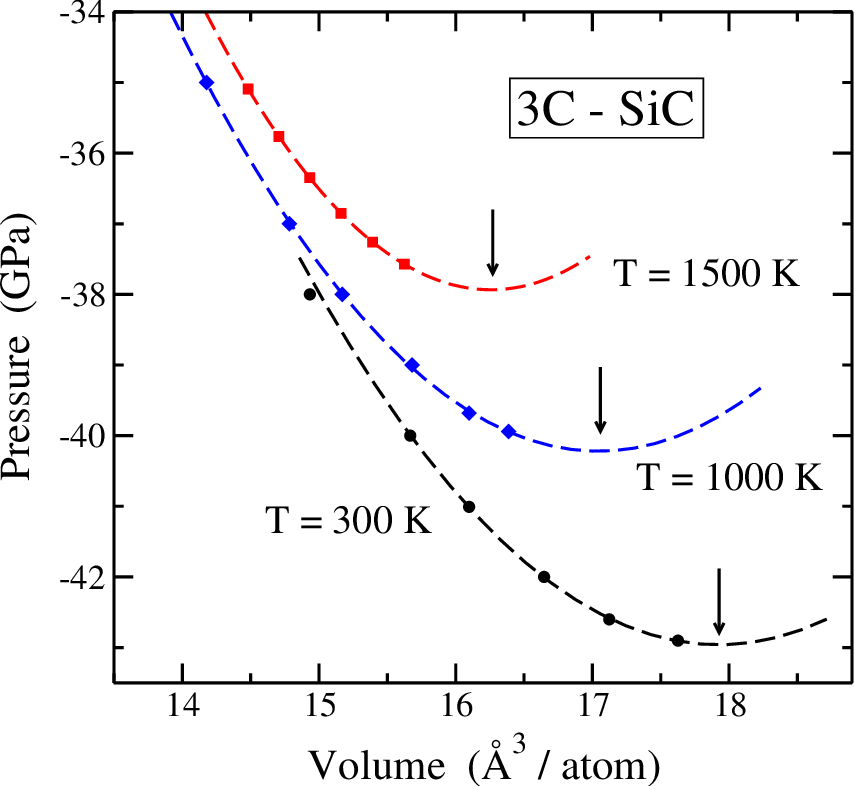}
\vspace{0.3cm}
\caption{Pressure vs volume for $3C$-SiC at $T$ = 300 K
(circles), 1000 K (diamonds), and 1500 K (squares),
derived from MD simulations.
Error bars are in the order of the symbol size.
The dashed lines are parabolic fits to the data points.
Arrows indicate the spinodal volume for each temperature.
}
\label{f5}
\end{figure}

To define the spinodal volume, $V_s$, and pressure, $P_s$, we have
carried out for each considered temperature a fit of our data close
to the spinodal instability to the expression 
$P = P_s + c (V_s - V)^2$ [see Eq.~(\ref{pfv}), with $c = 3 a_3$].
In Fig.~5 we show the fits corresponding to $T$ = 300, 1000, and 1500~K.
In each of these fits we considered the five data points nearest to
the instability. In this figure, arrows indicate the spinodal
volumes for the given temperatures.
Following this procedure, we have obtained $P_s$ and $V_s$ for several
temperatures in the range from 300 to 1500~K.

\begin{figure}
\vspace{-0.0cm}
\includegraphics[width=7cm]{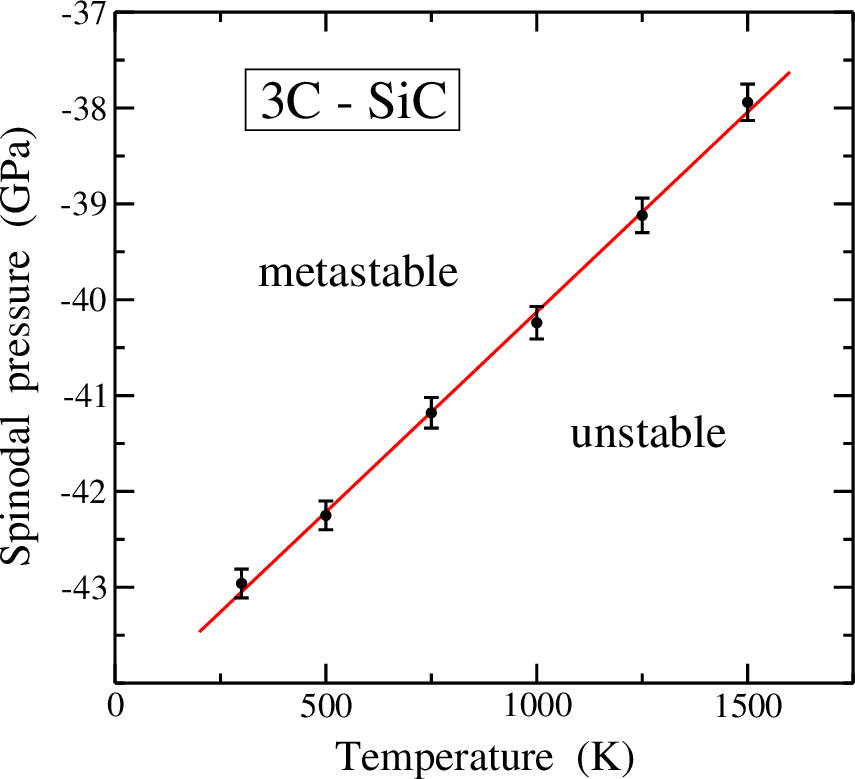}
\vspace{0.3cm}
\caption{Calculated spinodal pressure for $3C$-SiC
as a function of temperature.
Solid circles are data points derived from MD simulations.
The solid line is a least-square fit to the data points,
showing the separation between regions of metastability and
mechanical instability for $3C$-SiC.
}
\label{f6}
\end{figure}

In Fig.~6 we present the temperature dependence of the spinodal 
pressure of $3C$-SiC, as derived from our MD simulations
(solid symbols). The solid line is a fit to the data points:
$P_s = b_1 + b_2 T$, with  $b_1 = -44.3$~GPa and $b_2 = 4.2$~MPa/K.
The solid is metastable at negative pressures in the region 
above the line in Fig.~6. Below the line it is mechanically
unstable, so that in this region it transforms into the gas phase,
and the volume diverges to infinity under tensile pressure.
When approaching the line from the metastable region, the transition
may happen well before arriving at the spinodal, as occurs in our
isothermal-isobaric MD simulations when increasing the tensile 
stress or the temperature.

For comparison with the finite-temperature data obtained with
the TB model, we have calculated the spinodal pressure at $T = 0$
from DFT calculations. In this case, we obtain the
pressure as $P = - \partial E / \partial V$, and $P_s$ is given 
by the condition $\partial P / \partial V = 0$.
We find $P_s = -39$~GPa, which means that the tight-binding method
overestimates the spinodal pressure by about a 10\%.

Up to now, we have studied the dependence of the spinodal pressure 
on the temperature. Conversely, one can consider the temperature
at which the solid becomes unstable as a function of the crystal volume.
These results are summarized in Fig.~7, where we display the
spinodal temperature vs. the volume $V$.
Solid circles are data points derived from fits of the $P-V$
curves yielded by MD simulations to the expression in Eq.~(\ref{pfv}). 
Error bars in the volume are associated to the uncertainty in
the spinodal volume derived in the corresponding fits, as those
shown in Fig.~5.
A linear fit to the points in Fig.~7 yields a slope 
$d T_s / d V = -704$ K/\AA$^3$, and extrapolates at $T = 0$ to 
a volume $V = 18.4$~\AA$^3$/atom. This corresponds to a lattice
parameter $a$ = 5.28~\AA, consistent with calculations 
based on the zero-temperature energy curve shown in Fig.~1 
(dashed line), where the inflection point is indicated by 
a vertical arrow.

All the results presented here correspond to classical calculations 
and MD simulations. This means that nuclear quantum effects,
which should appear at low temperatures are not taken into account.
Thus, the low-$T$ limit, which is presented here as a reference
for finite temperature results correspond to the classical limit,
and does not take into account quantum corrections as those
arising from atomic zero-point motion.
An analysis of low-temperature quantum corrections to the results
presented here is out of the scope of the present paper, and could
be analyzed by means of path-integral simulations, as those 
employed earlier to study spinodal instabilities in noble-gas
solids.\cite{he03b}

\begin{figure}
\vspace{-0.0cm}
\includegraphics[width=7cm]{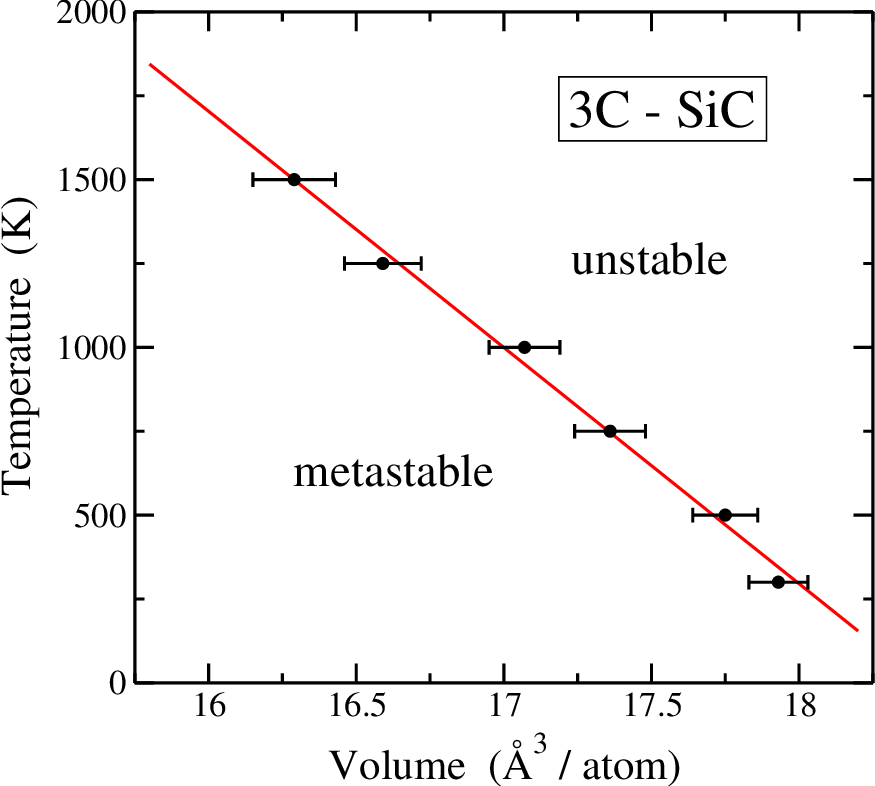}
\vspace{0.3cm}
\caption{Spinodal temperature vs volume for $3C$-SiC, as derived
from MD simulations (solid circles).
The line is a least-square fit to the data points.
In the regions below and above the line cubic silicon carbide
is metastable and mechanically unstable, respectively.
}
\label{f7}
\end{figure}

\section{Summary}

In this paper we have presented and discussed results of MD simulations
of cubic silicon carbide in a large range of temperature and pressure.
This method has permitted us to quantify structural and elastic
properties of this crystalline semiconductor, with particular emphasis 
upon its limit of mechanical stability.

We have concentrated on the elastic constants and the region of mechanical 
stability under tensile pressure. With this purpose,
we have put forth the results of extensive simulations of
this material using a well-checked tight-binding Hamiltonian, for
a wide range of temperatures and hydrostatic pressures. 
The results of our MD simulations have been found to be consistent 
with DFT calculations at $T = 0$ in a large range of crystal volumes 
and pressures.
This has served us as a check for the precision of the TB Hamiltonian
employed here to study silicon carbide for crystal volumes far
from the equilibrium state at ambient conditions.

For $P = 0$, the elastic constants $C_{11}$ and $C_{12}$ 
of cubic SiC, as well as the Poisson's ratio $\nu$, 
are found to decrease for rising temperature, as discussed in Sec.~IV.
This decrease is even more important in the presence of tensile stress,
so that at $T$ = 300~K, $C_{12}$ and $\nu$ become negative for a 
pressure $P = -27$~GPa ($3C$-SiC converts into an auxetic material).
For larger negative pressure, one reaches the spinodal instability,
where the solid becomes mechanically unstable (vanishing bulk modulus). 
For $T = 300$~K, this happens at $P_s = -43$~GPa, a spinodal pressure 
which is less negative for higher $T$: $P_s = -37.9$~GPa at 1500 K.

The computational approach presented in this paper has proven
to be a reliable tool to describe the effect of pressure in
metastable states in solids. 
In particular, it allows to determine the spinodal line
under tensile stress as a function of temperature.
Further work in this subject is necessary to generalize the
results presented here to other related crystalline materials, 
for which the stability limits are expected to depend on their 
elastic properties. This can be realized by means of atomistic 
simulations using accurate tight-binding Hamiltonians as that
employed here for SiC.   \\  \\

\noindent
{\bf Data availability} 

The data that support the findings of this study are available
from the corresponding author upon reasonable request.  \\

\noindent
{\bf CRediT author contribution statement} 

Carlos P. Herrero: Data curation, Investigation, Validation, Original draft

Rafael Ram\'irez: Methodology, Software, Investigation, Validation 

Gabriela Herrero-Saboya: Methodology, Investigation, Validation  \\

\noindent
{\bf Declaration of Competing Interest}  

The authors declare that they have no known competing financial
interests or personal relationships that could have appeared to
influence the work reported in this paper.  \\

\begin{acknowledgments}
This work was supported by Ministerio de Ciencia e Innovaci\'on
(Spain) through Grant PGC2018-096955-B-C44.
\end{acknowledgments}



\end{document}